\documentclass[sigconf,nonacm, 10pt]{acmart}
\usepackage{subcaption}
\usepackage{filecontents}
\usepackage{nameref}
\usepackage{tikz}
\usepackage{amsmath}

\usepackage{enumitem}
\usepackage{pbox}
\usepackage{multirow}
\usepackage{tabularx}
\usepackage{makecell}
\usepackage{mathtools}
\usepackage{caption}
\usepackage{float}
\usepackage{booktabs}
\usepackage[linesnumbered,ruled,vlined]{algorithm2e}
\usepackage{algorithm2e}
\usepackage{xcolor}
\usepackage{nameref}
\usepackage{wrapfig}
\usepackage{xspace}
\usepackage{url}
\usepackage{pifont}
\usepackage{xcolor}

\usepackage{breakurl}
\usepackage{hyperref}

\newcommand{\parab}[1]{\vspace{0.03in}\noindent\textbf{#1}}


\title[Towards an Agentic Workflow for Internet Measurement Research]{Towards an Agentic Workflow for \\ Internet Measurement Research}

\author{Alagappan Ramanathan}
\orcid{0009-0003-3293-1790}
\affiliation{%
   \institution{University of California, Irvine}
   \country{USA}
}
\author{Eunju Kang}
\orcid{0009-0005-3591-5758}
\affiliation{%
    \institution{University of California, Irvine}
    \country{USA}
}
\author{Dongsu Han}
\orcid{0000-0001-6922-7244}
\affiliation{%
  \institution{KAIST}
  \country{Republic of Korea}
}
\author{Sangeetha Abdu Jyothi}
\orcid{0009-0000-0503-4478}
\affiliation{%
   \institution{University of California, Irvine}
   \country{USA}
}
\thanks{Sangeetha Abdu Jyothi holds concurrent appointments at UC Irvine and Amazon. This publication describes work performed at UC Irvine and is not associated with Amazon}

\renewcommand{\shortauthors}{Alagappan Ramanathan, Eunju Kang, Dongsu Han, Sangeetha Abdu Jyothi}

\begin{abstract}
Internet measurement research faces an accessibility crisis: complex analyses require custom integration of multiple specialized tools that demands specialized domain expertise. When network disruptions occur, operators need rapid diagnostic workflows spanning infrastructure mapping, routing analysis, and dependency modeling. However, developing these workflows requires specialized knowledge and significant manual effort. 

We present ArachNet, the first system demonstrating that LLM agents can independently generate measurement workflows that mimics expert reasoning. Our core insight is that measurement expertise follows predictable compositional patterns that can be systematically automated. ArachNet operates through four specialized agents that mirror expert workflow, from problem decomposition to solution implementation.
We validate ArachNet with progressively challenging Internet resilience scenarios. The system independently generates workflows that match expert-level reasoning and produce analytical outputs similar to specialist solutions. Generated workflows handle complex multi-framework integration that traditionally requires days of manual coordination. ArachNet lowers barriers to measurement workflow composition by automating the systematic reasoning process that experts use, enabling broader access to sophisticated measurement capabilities while maintaining the technical rigor required for research-quality analysis.
\end{abstract}

\begin{document}


\maketitle

\section{Introduction}

Internet measurement research faces a significant accessibility challenge. Complex analyses require orchestrating multiple specialized tools---BGP analyzers~\cite{routeviews, riperis, bgptools, bgpstream, bgpdetecting, bgpstate, bgpnext, albgp}, traceroute processors~\cite{tracerouteinternet, traceroutemao, paristraceroute, traceroutemultilevel}, topology mappers~\cite{topologyzoo, nautilus, topologycollecting}, and performance monitors ~\cite{perfsurvey, gatechIODA, netblocks, xaminer, cloudflare}---each with unique interfaces, data formats, and domain knowledge requirements. When researchers need to understand routing behavior, infrastructure dependencies, or performance anomalies, they must manually integrate different measurement systems through custom solutions. This creates a substantial barrier: the ability to compose advanced measurement workflows requires specialized domain experience, limiting such capabilities to a small community of experts.

Recent events highlight this challenge's practical impact. The 2022 AAE-1 cable cuts~\cite{multiple_providers_aae1_outage} and FALCON cable failure~\cite{falcon_outage} caused widespread outages, requiring rapid development of workflows integrating cable mapping, BGP analysis, and traffic flow assessment. Similar challenges arise regularly: CDN performance degradation requires correlating traceroute data with BGP changes~\cite{latlong, cdnmoving, cdnserver}; security incidents need workflows integrating multiple measurement perspectives. In Internet resilience research, while recent measurement frameworks~\cite{nautilus,xaminer} offer powerful capabilities, these tools operate in isolation and require specialized knowledge. Each scenario requires experts spending days developing measurement workflows before beginning analysis. What remains absent is a general, flexible measurement framework accessible beyond a narrow group of experts.

We take an alternative view. What if network operators could ask, "How would losing the Europe-Asia cables affect major content providers?" and receive executable measurement workflows in minutes? What if researchers could compose Internet measurement tools without specialized training in each framework? Today, this seems impossible.

We present ArachNet, the first system to demonstrate that LLM agents can independently generate measurement workflows that capture expert reasoning patterns. Our key contribution is recognizing that measurement workflow development follows predictable patterns, with expert reasoning broken down into distinct phases: problem analysis, solution design, implementation, and adaptation. ArachNet executes these phases through four specialized agents operating on a curated registry that captures measurement tool capabilities through standardized representations. 

ArachNet's coordinated pipeline works as follows: QueryMind breaks down problems into sub-problems, identifies dependencies, and estimates constraints and risks. Building on this analysis, WorkflowScout transforms these sub-problems into solution workflows by systematically exploring optimal combinations of registry functions. SolutionWeaver then converts the workflow design into executable code that users run to solve their measurement problems. Finally, RegistryCurator identifies useful capabilities from successful solutions and adds them to the registry for future use. ArachNet focuses specifically on workflow composition and code generation, and \textit{not} on improving individual measurement tools or data collection. Users express goals in natural language, and the system automatically generates executable measurement solutions that provide complete workflows or serve as foundations for expert refinement.

To validate these capabilities, we focus on Internet resilience analysis---a domain that exemplifies the workflow composition challenge by requiring integration across infrastructure mapping, routing analysis, and dependency modeling. Our evaluation demonstrates ArachNet's ability to \textit{(i)} independently generate workflows that produce analytical outputs similar to expert-designed solutions, \textit{(ii)} orchestrate cascading failure analysis across multiple measurement frameworks with significant integration complexity, and \textit{(iii)} perform temporal forensic investigations that match domain specialist approaches in methodology and results. These scenarios provide promising tests of expert-level reasoning because they require the same architectural decisions and tool integration strategies that specialists use in practice.

ArachNet lowers barriers to measurement workflow composition by automating the systematic reasoning process that experts use. New researchers can now tackle sophisticated analyses without deep specialization in each tool. During critical incidents, teams can rapidly compose diagnostic solutions spanning multiple measurement domains. Meanwhile, experienced researchers gain a force multiplier, focusing on novel insights while ArachNet handles integration complexity. We open source ArachNet's prompts and the case studies~\footnote{Prompts and case studies available at \href{https://gitlab.com/netsail-uci/arachnet}{https://gitlab.com/netsail-uci/arachnet}}.

\section{Related Work}

There has been research using agentic AI with LLM in networking problems such as design, configuration, and diagnosis. ChatNet~\cite{chatnet} handles networking tasks from natural language queries but still depends on human intervention. NADA~\cite{nadaHotnet2024} uses LLMs to generate network algorithms, though generated designs require quality checks. Zhou et al.~\cite{agentresearcherHotnet2023} propose an LLM-based agent that retrieves knowledge from Web resources and iteratively self-learns, but generating high-quality research questions still requires human evaluation expertise. Kotaru~\cite{operatorHotnet2023} applies LLMs to help operators translate natural language queries into metric-driven code, yet integration remains challenging due to inconsistent data formats. Unlike these approaches, ArachNet enables end-to-end automated workflow composition across different measurement tools through a multi-agent architecture that systematically captures expert reasoning patterns from problem decomposition to executable implementation.
\section{Design}

\begin{figure*}
    \centering
    \includegraphics[width=\linewidth]{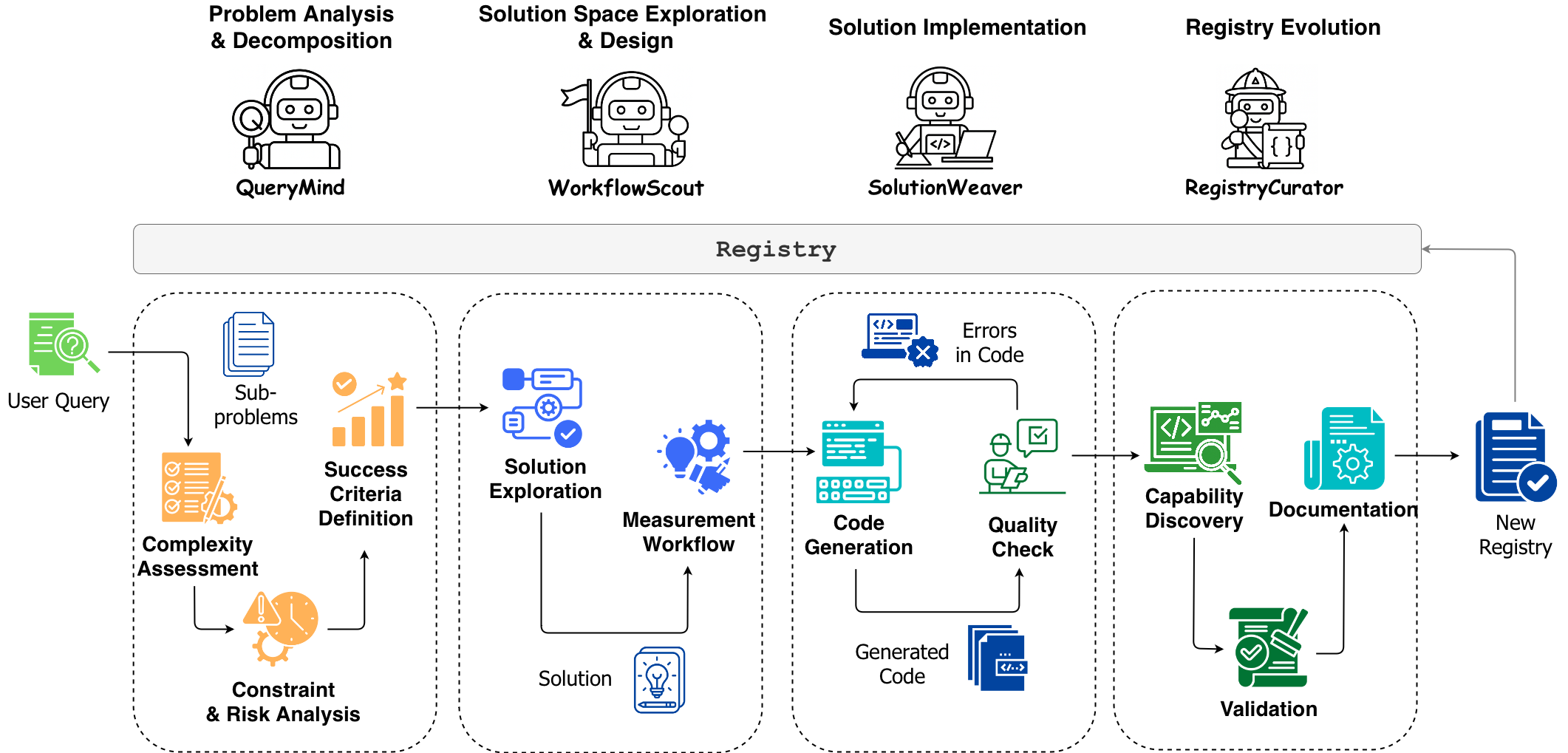}
    \caption{\small ArachNet design with four specialized agents. QueryMind analyzes and decomposes problems, WorkflowScout explores and designs solution workflows, SolutionWeaver implements executable workflows, and RegistryCurator evolves ArachNet's capabilities. }
    \label{fig:design_diagram}
\end{figure*}

ArachNet treats Internet measurement as a compositional problem where complex analyses emerge from intelligently combining expert-built building blocks. Unlike existing manual frameworks requiring users to know which tools exist and how to wire them together, ArachNet captures the problem-solving process itself—the systematic approach experts use to navigate from query to solution. ArachNet uses four specialized agents with carefully designed prompts that mirror how experts work (Figure~\ref{fig:design_diagram}) with each agent handling a distinct reasoning phase. By default, the system runs in "standard" mode for fully automated workflows. In "expert" mode, domain specialists can review and adjust outputs between agents before proceeding to the next stage.

\parab{Registry: Measurement Capability Encoding.} The Registry forms ArachNet's foundation—a manually curated catalog describing what measurement tools can do, not how they do it. This design emerged from early experiments: exposing entire codebases to agents overwhelmed them with implementation details, causing them to miss key capabilities buried in thousands of lines of code. ArachNet’s compact registry instead provides agents with a "measurement API" for intelligent composition. Each entry specifies a tool's capabilities (e.g., "maps IP links to submarine cables"), required inputs, expected outputs, and constraints. This scales linearly with available tools, transforming diverse measurement frameworks into unified knowledge that supports automated reasoning.

\parab{QueryMind: Problem Analysis \& Decomposition.} QueryMind transforms user queries into structured sub-problems with clear dependencies and constraints. This separation exists because problem understanding requires different reasoning than solution design—experts first clarify what needs to be measured before considering how to measure it. The agent recognizes that seemingly simple queries contain hidden complexity that experts instinctively identify. A request like "measure CDN performance" actually includes latency analysis across regions, cache behavior evaluation, and temporal consistency checking. QueryMind agent systematically breaks queries into manageable pieces, examining data dependencies, causal relationships, spatial (geographic) constraints, and temporal aspects to classify problems and guide solution strategy.

Building on this structured decomposition, QueryMind evaluates constraints that fundamentally shape feasible solutions. The agent analyzes data availability, technical (registry tools) constraints, and methodological limitations early in the process because constraints determine what approaches are feasible. If required data is unavailable or technical limits exist, certain solutions become impossible regardless of their merit. This constraint analysis reveals potential measurement gaps and failure modes that could compromise results. QueryMind then defines success criteria to determine when queries are sufficiently answered. Without explicit success criteria, agents risk both under-analysis (stopping too early) and over-engineering (continuing unnecessarily). In expert mode, specialists refine this understanding by adjusting scope, adding constraints, or highlighting dependencies.

\parab{WorkflowScout: Solution Space Exploration \& Design.} WorkflowScout agent converts QueryMind's structured sub-problems into concrete solution architectures. We separate exploration from implementation because solution design requires evaluating trade-offs across multiple approaches, fundamentally different from writing code that focus on quality checks and error handling to execute a chosen approach.

The agent systematically explores available capabilities using an adaptive strategy that scales exploration effort to match problem complexity. For simple queries (e.g., single-framework analyses), WorkflowScout evaluates a direct solution path since exploring alternatives provides minimal benefit. For complex queries requiring multi-framework integration, it explores multiple approaches and compares trade-offs in data requirements, computational complexity, and reliability. This selective evaluation matters because measurement problems often have multiple valid approaches with different trade-offs only apparent through systematic comparison. WorkflowScout then designs optimal combinations of registry functions by analyzing data flow patterns, temporal requirements, and validation opportunities, ensuring coherent end-to-end workflows by resolving data dependencies and optimizing execution order. In expert mode, specialists guide exploration by suggesting alternative approaches or imposing additional quality constraints.

\parab{SolutionWeaver: Solution Implementation.} SolutionWeaver agent converts solution architectures into executable code integrating heterogeneous measurement tools. The agent tackles integrating independently designed measurement tools. It implements format translation using registry specifications to ensure seamless data flow, converting BGP data formats to match topology mapper inputs, or transforming traceroute outputs for statistical analysis tools. This translation layer is critical because measurement tools use diverse data representations creating integration barriers. Quality assurance is woven throughout the implementation rather than added post-hoc. SolutionWeaver embeds automated checks during code generation, covering consistency verification across data sources, sanity checking of measurement results, and uncertainty quantification. This generates quality metrics that help users interpret results appropriately. In expert mode, specialists review and optimize generated code based on deployment experience.

\parab{RegistryCurator: Systematic Registry Evolution.} As workflows are built and run successfully, patterns emerge that could be useful for future queries. RegistryCurator ensures ArachNet's capabilities grow organically by identifying reusable patterns from successful workflows and adding them to the registry. This agent exists because manual registry curation doesn't scale—as the system generates more workflows, human experts cannot feasibly review every potential capability. The agent analyzes successful workflow implementations to identify reusable patterns across data processing utilities, analysis algorithms, and integration functions. Validation happens before integration to prevent registry bloat—only capabilities proving useful in practice merit inclusion rather than speculatively adding every possible function. This validation-first strategy ensures that not all workflow patterns are worth generalizing; only those demonstrating both accuracy and utility across multiple use cases should be added. Once validated, RegistryCurator generates the structured representation and documentation needed for registry integration. In expert mode, specialists validate that identified capabilities meet community standards and maintain consistency with existing registry conventions.

In summary, ArachNet's four-agent architecture enables users to express measurement goals in natural language and generate executable solutions. By automating the systematic reasoning process that experts use—problem decomposition, constraint analysis, solution exploration, and implementation—ArachNet lowers barriers to measurement workflow composition while maintaining flexibility for domain specialists to guide and refine outputs through expert mode.

\section{Case Studies}

We develop an ArachNet prototype using specialized prompts for each of the four agents based on Claude Sonnet 4. These prompts evolved through iterative refinement—when generated outputs faltered (missing constraints, overlooking dependencies, or proposing unnecessarily complex solutions), we embedded the generalized reasoning a human expert would naturally apply into the prompt. The core reasoning patterns proved transferable across measurement problems, though domain-specific knowledge required careful encoding in both registry entries and agent prompts. To use ArachNet, users provide natural language queries and ArachNet generates executable Python code that users run to solve their measurement problems. To demonstrate ArachNet's effectiveness, we present validation through progressively challenging scenarios. We start with expert solution replication and advance to novel multi-framework integration. Although ArachNet is designed for general Internet measurement queries across diverse domains, we focus our case studies on Internet resilience frameworks. This provides deep validation of our approach within a well-defined measurement domain.

\subsection{Level 1: Expert Solution Replication}

Can ArachNet independently arrive at solutions equivalent to those generated by domain experts? We validate this functionality by comparing generated workflows against expert implementations. We use the Xaminer framework as our benchmark, which performs cross-layer resilience analysis using the mapping results generated by Nautilus framework.

\vspace{2mm}
\parab{\underline{Case Study 1}: Expert-Level Cable Impact Analysis}

\parab{Challenge:} ``Identify the impact at a country level due to SeaMeWe-5 cable failure''

\parab{Why This Is Hard:} This seemingly simple query requires expert-level decomposition. The system must understand cable dependencies, extract affected IP addresses, perform geographic mapping, and aggregate country-level impacts. Traditional approaches require deep knowledge and manual integration of several frameworks and data transformations.

\parab{Setup:} We provide the agent with only core Nautilus system functions. We withhold Xaminer's higher-level abstractions to test whether equivalent workflows can be independently derived. This controlled setup ensures the agent relies purely on analytical reasoning rather than following guided architectural patterns.

\parab{Summary:} ArachNet closely follows Xaminer workflows with significant functional overlap. It achieves equivalent country-level impact analysis without domain-specific architectural guidance using $\approx$ 250 lines of code. 

\parab{Detailed Technical Comparison:} Xaminer uses sophisticated embedding modules that aggregate cross-layer metrics at country and AS-level abstractions through normalized metrics including IPs, links, ASes, and AS links per country. In contrast, ArachNet independently develops a direct processing pipeline. This pipeline systematically transforms mapping data from Nautilus into country-level assessments. Despite this architectural difference, ArachNet's approach produces similar impact metrics. It generates geographic distribution analysis that provides enhanced analytical insights.

Both workflows achieve functionally equivalent logic. This includes cable dependency identification, IP extraction, geographic mapping, and country-level aggregation. The agent identifies these same essential data transformations as expert-designed Xaminer. It does this without prior architectural knowledge, demonstrating that complex measurement reasoning can be systematically automated.

\vspace{3mm}
\parab{\underline{Case Study 2}: Natural Disaster Impact Analysis}
\parab{Challenge:} ``Identify the impact of severe earthquakes and hurricanes globally assuming a 10\% infra failure probability''

\parab{Why This Is Hard:} Multi-disaster analysis requires complex cross-framework integration given the diversity of disaster types and thresholds. The key challenge is determining whether sophisticated multi-system orchestration is necessary or if simpler approaches suffice. This decision requires proficient architectural judgment to avoid both under-engineering and over-engineering solutions.

\parab{Setup:} We provide registry functions from multiple frameworks to test the agent's decision-making. We want to see whether the agent will identify that Xaminer's single event processing capability alone can handle multi-disaster analysis and will avoid unnecessary cross-framework integration.

\parab{Summary:} ArachNet demonstrates skilled restraint by correctly identifying that complex multi-disaster analysis requires only a single function. It avoids unnecessary over-engineering with only $\approx$ 300 lines of code, even when presented with multiple available tools.

\parab{Detailed Technical Comparison:} Both ArachNet and Xaminer workflows are functionally identical. They leverage the event processing function's versatility to handle earthquakes and hurricanes separately. The workflows apply failure probabilities and combine results for comprehensive global impact metrics through the same computational approach. Critically, ArachNet avoids incorporating functions from other available frameworks when a single function provides all necessary capabilities. This reflects expert-level solution scoping based on actual requirements rather than available capabilities. It demonstrates sophisticated architectural judgment that matches domain expert decision-making.

\subsection{Level 2: Multi-Framework Orchestration}

Having validated expert-level reasoning on single-framework problems, we now examine whether ArachNet can handle complex scenarios requiring integration across multiple measurement systems. These cases demonstrate capabilities that push beyond current tool limitations. They enable analyses that were previously impractical due to the expertise and time required for multi-system coordination.

\vspace{3mm}
\parab{\underline{Case Study 3}: Automated Cascading Failure Analysis}

\parab{Challenge:} ``Analyze the cascading effects of submarine cable failures between Europe and Asia''

\parab{Why This Is Hard:} Cascading failure analysis requires sophisticated integration across multiple measurement domains. This includes infrastructure mapping, impact analysis, temporal correlation, and cross-layer synthesis. Traditional approaches require researchers to separately run cross-layer systems for mapping and analysis. They must also use BGP and traceroute tools for temporal analysis, then manually correlate results through custom scripts and domain expertise. This process requires deep knowledge of multiple frameworks and their limitations. It often takes days to properly integrate measurement systems for comprehensive analysis.

\parab{Summary:} ArachNet automates integration across 4 frameworks spanning infrastructure, topology, and temporal domains. It orchestrates analysis comprising $\approx$ 525 lines of code, that traditionally requires days of manual coordination into seamless automated workflows.

\parab{Detailed Investigation Results:} The agent demonstrates sophisticated cross-framework integration across multiple domains. Primary integration combines Nautilus cable mappings with Xaminer impact analysis, automatically transforming data formats and implementing geographic filtering to focus on Europe-Asia connectivity. Building on this foundation, secondary integration leverages submarine cable and AS dependency graphs for cascade propagation modeling using graph algorithms that trace failure propagation paths. Temporal integration then combines BGP dumps and traceroute data for evolution analysis. This tracks how failures manifest over time across different measurement perspectives. Most significantly, the agent implements cross-layer synthesis that integrates all outputs into unified cascade timelines. These timelines span cable, IP, and AS layers, providing comprehensive failure analysis. This would typically require extensive manual coordination across tools. Our agentic approach automatically identifies these integration requirements and orchestrates multi-framework workflows without manual intervention. This demonstrates the potential to significantly accelerate complex measurement research that currently represents major bottlenecks in the field.

\subsection{Level 3: Forensic Analysis}

A sophisticated test of ArachNet's capabilities involves expert-level analysis requiring temporal correlation and causation establishment. This represents a pinnacle in measurement expertise. It requires integration of statistical analysis, infrastructure knowledge, and routing behavior understanding to establish definitive causal relationships.

\vspace{3mm}
\parab{\underline{Case Study 4}: Automated Root Cause Investigation}

\parab{Challenge:} ``A sudden increase in latency was observed from European probes to Asian destinations starting three days ago. Determine if a submarine cable failure caused this, and if so, identify the specific cable.''

\parab{Why This Is Hard:} This temporal forensic scenario requires integration of traceroute measurements, BGP routing data, and cable infrastructure mappings across a specific time window. The goal is to establish causation between network events and observed anomalies. Traditional forensic analysis requires expert knowledge across multiple domains including statistical anomaly detection, infrastructure correlation, routing analysis, and evidence synthesis. This typically requires extensive manual analysis across multiple measurement systems. Experts spend days or weeks correlating evidence from different sources to establish definitive causation.

\parab{Summary:} ArachNet successfully implements temporal correlation algorithms with causation establishment and definitive cable identification with $\approx$ 750 lines of code. It demonstrates advanced forensic capabilities that eliminate traditional manual analysis bottlenecks while maintaining rigorous evidence standards.

\parab{Detailed Investigation Results:} The agent implements comprehensive forensic analysis across multiple analytical domains with systematic evidence integration. Statistical analysis implements anomaly detection on traceroute latency data. It establishes quantitative baselines and detects significant increases with proper significance assessment to ensure robust anomaly identification. The system then uses cable mapping data to identify which submarine cables might be responsible. It applies scoring algorithms to rank each cable by likelihood of involvement. Complementing this infrastructure analysis, BGP validation processes routing dumps to detect temporal correlation between routing changes and latency anomalies. This provides independent verification of infrastructure-level hypotheses. Finally, the system combines evidence from all three analyses. Statistical analysis provides the anomaly detection, infrastructure correlation identifies suspect cables, and routing validation confirms the timing. This comprehensive approach provides confidence scores and identifies the specific failed cable. The system establishes clear causation between the cable failure and observed latency. Traditional analysis would require days of manual work across multiple tools. Our system automates this forensic process while matching expert-level reasoning.

\parab{Summary.} Our evaluation demonstrates that ArachNet successfully addresses the core challenges in Internet measurement research. First, it captures and applies expert-level reasoning without domain-specific guidance, matching sophisticated analytical workflows developed by measurement specialists. Second, it automatically orchestrates complex multi-framework analyses that traditionally require days of manual integration effort. Finally, it performs advanced forensic investigations with systematic evidence correlation that eliminates traditional analysis bottlenecks. These results validate our core thesis: agentic systems can democratize complex Internet measurement capabilities while preserving the technical rigor required for research-quality analysis.

\section{Research Challenges}

While ArachNet demonstrates the feasibility of automated measurement workflow composition, several challenges and opportunities emerge from our work that point toward important future research directions.

\parab{Generated Code Quality and Domain Knowledge Capture: } Our evaluation reveals an important difference between domain-specific reasoning and regular programming implementation. While generated workflows sometimes contain minor coding errors, the system successfully captures and applies complex Internet measurement domain knowledge. ArachNet shows sophisticated understanding of measurement tool capabilities, appropriate integration patterns, and analytical reasoning that typically requires specialized measurement expertise. The remaining errors are standard programming issues that do not require specialized knowledge to fix, suggesting that the core challenge of domain expertise transfer has been successfully addressed while leaving manageable implementation improvements.

\parab{Prompt Engineering and Generalization: } An important open question is how ArachNet's approach generalizes to new measurement domains beyond Internet measurements or to different LLM architectures. While our case studies demonstrate effectiveness within a focused domain, systematic investigation is needed to understand adaptation requirements for diverse scenarios such as application performance analysis, security monitoring, or network operations. Future work should investigate methods for reducing domain-specific prompt engineering effort, techniques for making the system less dependent on specific LLM capabilities, and approaches for validating that core reasoning patterns transfer reliably across disciplines.

\parab{Trust and Verification: } A critical challenge for automated workflow generation is establishing trust in system outputs. While our case studies demonstrate functional equivalence to expert solutions in specific scenarios, several verification questions remain open. How do we validate that generated workflows are correct for novel queries without expert ground truth? What guarantees can we provide about workflow correctness?

Unlike domains such as algorithm optimization where automated verifiers can objectively measure performance, measurement workflows present a fundamental verification challenge: correctness depends on methodological soundness, appropriate tool selection, and valid integration patterns—aspects that currently require expert judgment rather than automated evaluation. Ensemble methods comparing multiple independent workflow generations could provide confidence scores by identifying consensus approaches, while formal verification techniques might detect certain classes of logical errors (e.g., data type mismatches, missing dependencies). However, the core challenge of verifying that a workflow uses the right measurement methodology for a given query remains open. Developing mechanisms that can assess methodological validity, provide interpretable explanations of architectural decisions, and flag potentially problematic approaches will be crucial for building user trust and enabling broader adoption amongst non-experts.

\parab{Handling Conflicting Tool Outputs: } Real measurement scenarios often involve tools that provide contradictory results or work under different assumptions. For instance, BGP routing tables might show one path while traceroute reveals actual packet travel through different routes, or topology mappers may disagree on infrastructure connections. Future systems need smart conflict resolution methods that can detect inconsistencies, weigh tool reliability based on past accuracy, and generate workflows that gracefully handle disagreements through confidence scoring systems or meta-analysis approaches.

\parab{Seamless Research Workflow Integration: } Most researchers have established pipelines and preferred tools that cannot be easily replaced. Rather than requiring complete workflow replacement, future systems should support gradual adoption where AI-generated components can be smoothly integrated into existing pipelines while automating the entire execution process. This includes developing adapters for popular analysis frameworks, supporting hybrid workflows where some components are manually specified while others are automatically generated, and creating automated execution environments that handle deployment, dependency management, and result collection. Such systems would provide migration paths allowing researchers to gradually transition from manual workflow composition to end-to-end automation---from natural language queries to final results---while maintaining compatibility with their existing tools and practices. 

\parab{AI Agent Communication Protocol Intergration: } The emergence of agent communication protocols, such as Model Context Protocol (MCP) and Agent-to-Agent protocol (A2A), presents significant opportunities for standardizing how AI agents interact with measurement tools. MCP's server-client design could provide a unified interface for AI agents to interact with external measurement tools, dramatically simplifying registry maintenance and tool integration through automatic capability discovery and standardized interaction patterns. Additionally, A2A protocols could formalize communication between ArachNet's specialized agents, enabling more robust task delegation, state management, and coordination as sub-problems flow through the pipeline. Adopting standardized protocols could transform ArachNet from a custom implementation into a standards-based system where agent-to-tool and agent-to-agent interactions follow community-wide conventions. However, realizing these benefits requires widespread protocol adoption across both the measurement tool community and the AI agent ecosystem.

\parab{Scalability and Registry Evolution: } A key challenge is keeping the Arachnet registry accurate as measurement tools evolve. Our registry currently needs manual updates to capture tool capabilities, interfaces, and requirements. Specialized LLM agents could automatically analyze codebases and generate accurate registry descriptions. Such agents could continuously monitor tool repositories, read API documentation and release notes, and extract capability details, ensuring the registry stays current with minimal manual work.

The path forward requires addressing these challenges while preserving the core insight that expert reasoning patterns can be systematically captured and automated. As measurement tools become more sophisticated and research questions more complex, the need for intelligent workflow composition will only grow, making agentic frameworks critical for advancing measurement research.

\section*{Acknowledgment}

This work was supported in part by  Institute of Information \& Communications Technology Planning \& Evaluation (IITP) of the Korea government (MSIT) (No.\,RS-2024-00398157 and RS-2024-00418784). Sangeetha Abdu Jyothi and Dongsu Han are corresponding authors. 

\bibliographystyle{ACM-Reference-Format} 
\bibliography{paper}

\end{document}